\def\year{2017}\relax
\documentclass[letterpaper]{article}
\usepackage{aaai17nC}
\usepackage{times}
\usepackage{helvet}
\usepackage{courier}
\usepackage[noend]{algorithmic}[1]
\usepackage{amssymb}
\usepackage{amsmath}
\begin{document}

\renewcommand{\algorithmicrequire}{\textbf{Input:}}
\renewcommand{\algorithmicensure}{\textbf{Output:}}

\newcommand{\cH}{{\cal H}}
\newcommand{\cS}{{\cal S}}
\newcommand{\cA}{{\cal A}}

\title{Eqilibrium Approximation Quality of Current No-Limit Poker Bots}
\author{Viliam Lis\'{y}$^{a,b}$\\
$^a$Artificial intelligence Center\\
Department of Computer Science, FEL\\
Czech Technical University in Prague\\
viliam.lisy@agents.fel.cvut.cz\\
\And
Michael Bowling$^b$\\
$^b$Computer Poker Research Group\\
Alberta Machine Intelligence Institute\\
Dept. of Computing Science, University of Alberta\\
mbowling@ualberta.ca
}
\maketitle
\begin{abstract}
Approximating a Nash equilibrium is currently the best performing approach for creating poker-playing programs. While for the simplest variants of the game, it is possible to evaluate the quality of the approximation by computing the value of the best response strategy, this is currently not computationally feasible for larger variants of the game, such as heads-up no-limit Texas hold'em. In this paper, we present a simple and computationally inexpensive Local Best Response method for computing an approximate lower bound on the value of the best response strategy. Using this method, we show that existing poker-playing programs, based on solving abstract games, are remarkably poor Nash equilibrium approximations.
\end{abstract}

\noindent 
One very popular measure for progress in artificial intelligence is computers' performance in recreational games commonly played by humans.  There has been a dramatic sequence of successes in the past two decades starting with Chinook with checkers \cite{schaeffer1996chinook}, Deep Blue with chess \cite{campbell2002deep}, Watson with Jeopardy! \cite{ferrucci2013watson}, and AlphaGo with go \cite{silver2016mastering}.  Despite these successes, poker has proven to be a harder challenge for AI.  Similar to checkers, chess, and go, poker can be easily and completely described by a simple set of rules. The size of the game ranges from being smaller than checkers (heads-up limit Texas hold'em; HULH) to as large as go (heads-up no-limit Texas hold'em; HUNL).  However, it is a substantially more complex game due to the element of imperfect information. In poker, some information is private to specific players. Players need to infer the private information of others based on their actions in the game, while also seeking to avoid losing their own strategic advantage by revealing their private information via their actions.  This is complicated by the fact that the information revealed by an opponents' play depends on their behaviour, and that behaviour naturally depends on the player's own private information. 

Like other popular games, poker has been a challenge problem in artificial intelligence since the inception of the field \cite{kuhn1950simplified,koller1997representations,billings2002challenge}.  Recent progress in approximating Nash equilibria in massive extensive-form games \cite{gilpin2007gradient,CFR,lanctot2009monte,tammelin2015solving} has allowed for some initial successes in the smallest variant of poker played by humans, HULH.  Polaris, a program built around the CFR-family of methods, defeated professional poker players for the first time in a meaningful match of HULH in 2008 \cite{rehmeyer2008ante}.  In 2015, this was taken a step further by the program Cepheus, which essentially solved the game the HULH, with the resulting strategy requiring 11TB of storage and using over 900 CPU years of computation \cite{bowling2015heads}.  

The size of the game of HULH, though, is trivial compared to the more popular no-limit variants of the game.  In limit poker variants, there are at most 3 different actions available to a player at any situation (fold, check/call, bet/raise) as all bet amounts are fixed in advance.  In no-limit games, players can bet any number of remaining chips, leading to thousands of possible actions from a single situation.  As a result, HUNL can have over $10^{160}$ decision points in the game, as in the case of the variant played in the Annual Computer Poker Competition (ACPC)~\footnote{www.computerpokercompetition.org}.  This makes HULH's size of $10^{14}$ decision points seem trivial.  

Much of the progress in HULH was enabled by the ability to measure the approximation quality of a strategy in HULH~\cite{johanson2011accelerating}.  This has been a challenge for research in HUNL where to date evaluation has been limited to tournament evaluation,where strategies are evaluated by having them play against each other.  The results of such a tournament are necessarily relative and cannot give an absolute strength of any particular program.  Furthermore, it can substantially depend on small details in the design of the tournament, such as the winnings cap used in the ACPC~\cite{bard2016online}.  The absolute measure of performance through computing a strategy's approximation quality made possible a number of important strides in HULH research, for example, investigating the effect of restricted opponent modelling, asymmetric abstractions \cite{bard2014asymmetric}, translation, and payoff tilts \cite{johanson2011accelerating}.

This paper presents a simple method to quickly approximate a lower-bound to a strategy's exploitability in the HUNL game.  It is trivial to parallelize, makes no card abstraction commitments, can be applied even to strategies that use dynamic (and expensive) endgame solving techniques, and can probe a far larger portion of the total betting space than any current techniques use in their solving approach.  Using this technique, we compare a number of the top HUNL programs from the ACPC.  We show that even though the differences among the top performing agents in the ACPC are tiny fractions of a blind per hand, the exploitability of the players is several whole blinds per hand. For every program tested, it would be far less exploitable to immediately fold every hand than to use even such a state-of-the-art startegy.  Furthermore, we can tease apart the source of the approximation error, observing that considerably more exploitability can be attributed to card abstraction as opposed to betting abstraction.

\section{Current No-Limit Poker Bots}
While we are aware there are poker programs (or bots) playing online in real money games, this paper focuses on bots submitted to the ACPC. These bots are developed by top research teams, use principled AI approaches, and the techniques they use are to large extent well documented.

\subsection{Heads-up No-Limit Texas Hold'em}
Heads-up no-limit Texas hold'em is a variant of poker played with a standard deck of 52 cards. At the beginning of each hand or \emph{game}, the first player enters a \emph{big blind} into the \emph{pot}; the second player enters half of that size, or the \emph{small blind}; and both players are then dealt two \emph{private cards}. The second player then starts the first round of betting. The players alternate in choosing to \emph{fold} -- ending the game and letting the opponent take the pot; \emph{call} -- matching the amount of chips entered by the opponent and ending the betting round; or \emph{raise} by $x$ -- adding $x$ more chips than the opponent to the pot. A raise of all remaining chips is called an \emph{all in} bet. After the first round, three board or \emph{public cards} are dealt face up, and the first player now starts an identical round of betting to the first round. In the third and fourth rounds, one additional public card is dealt and betting starts again with the first player. If none of the players folds before the end of round 4, the game enters \emph{show-down}: The private cards are revealed and the pot is won by the player that can compose the strongest hand of 5 cards using his 2 private and the 5 public cards. A \emph{match} consist of large number of games, in which the players alternate their positions as the first and the second player.

\subsection{Best ACPC Players}
The bots annually submitted to the ACPC include programs based on hand-crafted rules, learning systems trained on logs of past games, or advanced linear programming methods.  The bots which have seen the most success in HUNL all have the same basic structure.

The bots are based on creating a smaller abstract version of the game, approximating the equilibrium strategy in the abstract game, and executing this strategy in the original game using a translation method to map real game situations into the abstraction. The abstract games abstract card information (called information abstraction) based on clustering hands with similar strength and potential to improve after additional cards are dealt. The abstract games abstract betting information by restricting the available bets to a small handful, usually expressed as fractions of the current size of the pot. The most successful bots approximate the Nash equilibrium in the abstract games using some variant of the Counterfactual Regret Minimization algorithm \cite{CFR}.

While playing a hand, the bots find the abstract state that corresponds to (cluster that includes) the current state of the game represented by the exact private and public cards in the game. Furthermore, they have to map the real betting sequence in the game that can use any size bets to the ``most similar'' betting sequence represented in the abstraction. The abstract strategy is queried for the probability distribution over actions (pot fractions) included in the abstract game and these are then post-processed and played in the actual game, if they are applicable. There are many publications related to each step of this process in the AI literature.

All three top performing players in ACPC 2016 match the high level description above. Their Instant Runoff Competition Results are summarized in Table~\ref{tab:ranking}. Baby Tartanian8 won the competition, Slumbot lost on average 12 mBB/h in its matches with the winner and Act1 lost 17 mBB/h on average against the other two agents.

\begin{table}[t]
\begin{small}
\begin{tabular}{c|c|c}
Bot Name & Authors & Winnings (mBB/h) \\\hline
Baby Tartanian8 & Carnegie Mellon Uni. & ~0.0 $\pm$  0.0  \\
Slumbot & Eric Jackson & -11.88 $\pm$ 10.33  \\
Act1 & Unfold Poker & -16.82  $\pm$ 14.35  \\
\end{tabular}
\end{small}
\caption{ACPC 2016 results.}\label{tab:ranking}
\end{table}

\section{Local Best Response}

This section presents the local best response algorithm for fast approximation of a lower bound on the exploitability of no-limit poker strategies. We call the player that computes the best response ``LBR'', and its opponent ``the opponent''. The key concept in this algorithm is the probability that the opponent holds each of the possible private hands, which we call the opponent's \emph{range}. At the very beginning of the game, it is equally likely that the opponent holds any pair of private cards, which is not in conflict with the cards held by the player. The probabilities of actions performed by the opponent depend on the private hand she holds. Therefore, with access to the strategy of the opponent, we can use Bayes' rule to infer the exact probabilities that the opponent holds each of the private hands. It is important that there is no abstraction or other approximation needed to exactly represent these probabilities. Based on the range, local best response greedily approximates best response actions, assuming a simple heuristic for behavior in the future.

Let $\cH$ be the set of all possible \emph{private hands}. In HUNL, each private hand consists of two cards. Ignoring their ordering, $|\cH|=1326$. A player's range is a probability distribution over hands and we denote it $\pi: \cH \rightarrow [0,1]$. We denote $\cS$ the set of \emph{public states} in the game. Each public state consists of the board cards, the order in which they came, and the complete sequence of bets by both players up to some point in the game. The \emph{strategy} of a player is a probability distribution on actions (fold, call, all different bet sizes) from $\cA$, available to a player: $\sigma: \cS\times\cH\times\cA\rightarrow [0,1]$.

The algorithm does not compute a best response strategy explicitly, but uses its local approximation to directly play against the evaluated strategy. At the beginning of each hand, LBR initializes the opponent's range uniformly for all private hands that do not include the cards dealt to LBR. After each action $a$ of the opponent performed at a public state $s$, LBR updates the opponent's range using her strategy $\sigma$:
\[ \pi(h)=\pi(h)\cdot\sigma(s,h,a) \]
and normalizes the distribution to sum to one. 

\begin{figure}[t]
\textbf{LocalBR}($\pi$ - range$, s \in \cS, h_i \in \cH$)

\begin{algorithmic}[1]
\STATE $wp = \mathit{WpRollout}(h_i, \pi, s)$
\STATE $asked = pot_{-i}(s) - pot_i(s)$
\STATE $U(call) = wp\cdot pot(s)-(1-wp)\cdot asked$
\FOR{action $a$ in considered bets / raises}
  \STATE $\mathit{fp} = 0$
  \FOR{opponent's hands $h_{-i} \in \cH $}
    \STATE $\mathit{fp} = \mathit{fp} + \pi(h_{-i})\cdot \sigma(s,h_{-i},\text{fold})$
    \STATE $\pi'(h) = \pi(h)\cdot(1-\sigma(s,h_{-i},\text{fold}))$
  \ENDFOR
  \STATE normalize $\pi'$
  \STATE $wp = \mathit{WpRollout}(h_i, \pi', s)$
  \STATE $U(a) = \mathit{fp}\cdot pot(s) +$\\
   $+(1-\mathit{fp}){\cdot}\left(wp{\cdot}(pot(s) {+} a)-(1{-}wp){\cdot} (asked {+} a)\right)$
\ENDFOR
\IF{$\max_a U(a) > 0$}
  \RETURN $\arg\max_a U(a)$
\ELSE
  \RETURN $\text{fold}$
\ENDIF
\end{algorithmic}
\caption{The algorithm for approximating lower bound on strategy exploitability.}\label{alg:lbr}
\end{figure}

LBR chooses its action to maximize its expected utility, under the assumption that the game will be checked/called until the end, unless the opponent folds right after LBR's action. The pseudocode is given in Figure~\ref{alg:lbr}. First, it computes the probability of winning the current hand if the game continues until the show-down in function {\it WpRollout}. This function exhaustively deals all possible remaining board cards and computes the mean probability of winning with hand $h_i$ against the opponent's range $\pi$. The expected utility of actions is computed with respect to utility 0 for fold. On line 3, the utility of action {\it call} for LBR is computed as the chips currently in the pot in case LBR wins, and the negative of the money LBR has to add to the pot in order to continue playing if it loses. Afterward, the algorithm computes the expected utility for all other considered actions. These are typically defined as a fixed set of pot fractions, but they can be arbitrary. For each action, lines 6-8 compute the probability that the opponent will fold after LBR performs the action given the current range ($\mathit{fp}$) and the new range that would hold for the opponent in case she does not fold ($\pi'$). The expected utility of the action for LBR (line 11) is computed as getting the whole pot if the opponent folds, getting the pot and the size of the bet ($a$) if the opponent does not fold and LBR wins, and losing the chips asked for and added otherwise.

This algorithm computes an approximation of the best response, looking only one action ahead and assuming that the players will check until the end of the game after performing the action (and not folding). The main advantage it exploits is that it perfectly understands the cards it holds and the state of the game without any abstraction. It may be easily extended to longer look-ahead or more complicated heuristics for estimating the value of the remainder of the game, however, it would substantially increase its computational requirements and even this simple version is very effective, as we show in the following section.

\subsection{Computing Lower Bound on Exploitability}

Recall that LBR does not pre-compute the best response approximation, but rather directly uses it to evaluate an input strategy. The evaluation consists of playing a large number of regular poker hands. The cards are dealt randomly as in a regular game. Every time it is the opponent's turn to play an action, her strategy is queried for the right probability distribution and an action is sampled based on the actual private hand held by the opponent. Every time it is LBR's turn to play, it updates the opponent's range and selects the best action based on the LocalBR algorithm in Figure~\ref{alg:lbr} and its private hand. The estimate of the exploitability is the average number of chips LBR wins in these games.

LBR queries the opponent's strategy for each hand after each action it considers. Therefore, if we want to run LBR with $n$ different pot fractions, completing 1 hand with LBR generally requires at most $(n|\cH|+1)$ times more computation time than playing a regular match with the strategy. Furthermore, the most expensive computation currently used in advanced poker bots is endgame solving \cite{ganzfried2016reflections}, which solves for all possible hands in one computation. If this is the dominant part of the computation for a strategy, LBR evaluation requires approximately $(n+1)$ times the computation required for playing a hand. If it is possible to play a game within few minutes on a single computational node, it is most likely also feasible to get reliable LBR estimates on a cluster of these nodes.

An important property of this algorithm is that it computes a lower bound on the exploitability of strategies. Since LBR actually plays a legal poker strategy, it can never win more in expectation than the worst case opponent of a strategy. Similarly, using longer look-ahead or different heuristic evaluation, as suggested above, would still have this property. Therefore, a strategy with substantially better performance against LBR is likely to be closer to an equilibrium. 

\subsection{Sampled soft translation}

When mapping the solution of the abstract game to the actual game played, it may be convenient to use sampling \cite{schnizlein2009probabilistic,ganzfried2013trans}. As a result, it may be difficult to obtain the exact probability distribution over actions in a particular public state of the real game. The proposed local best response method can still work in this situation. We can approximate the actual distribution by averaging a larger number of samples of the strategy at the same state. If we use a new independent sample from the strategy to pick the actual action played by the opponent, LBR does not learn any extra information about the action played and therefore computes a lower bound on the exploitability of the strategy.

\subsection{Variance reduction}
Since the evaluation plays-out standard poker hands, we can use any of the previously developed variance reduction techniques \cite{white2009learning,davidson2013baseline,burch2017aivat} to reduce the number of hands required to produce statistically significant results. For the experiments presented in this paper, we used duplicate matches and imaginary observations of expected outcomes of all hands the opponent could hold for a given line of play, instead of just the actual hand she holds. These two techniques combine to reduce the size of the confidence intervals by roughly 20\% with the same number of matches.

\section{Experimental evaluation}

In this section, we show that the proposed LBR computation is a fast and effective method to compute a lower bound approximation on exploitability of no-limit poker strategies.  We present results from running LBR on simple chump strategies; bots created at University of Alberta for past ACPCs; the two of the top three bots from the ACPC in 2016; and a huge strategy with a very sparse betting abstraction, but no card abstraction. Most results in this section will be presented in milli-big-blinds per hand (mBB/h) or whole big blinds per hand (BB/h).
The evaluation is stochastic; hence, we also present 95\% confidence intervals. In order to understand the common magnitude of these values, a bot that always folds as the first action would loose 750 mBB/h. The results of the one-on-one matches of the best three players in ACPC 2016 were all decided by less than 24 mBB/h.

In addition to showing the efficiency of the LBR computation, our experimental results also show that a large portion of the exploitability uncovered by the tool is caused by card abstraction. We show that using bets outside of the opponent's betting abstraction does add to the bot's exploitability, but at a significantly less magnitude.

\begin{table}[t]
\centering
\begin{small}
\begin{tabular}{c|c|c|c|c}
Betting & Rounds & Call & $\frac{1}{2}$Call$\frac{1}{2}$Raise & Random\\\hline
fc & 1-4 & 0 &  ~~7.1 $\pm$ 0.5 & 15.7 $\pm$ 0.4 \\
fc & 3-4 & 0 & 16.2 $\pm$ 0.3 & 2.2 $\pm$ 0.7\\\hline
fcpa & 1-4 & 34.0 $\pm$ 0.5 & 23.1 $\pm$ 0.5 & 39.1 $\pm$ 0.6\\
fcpa & 3-4 & 49.0 $\pm$ 0.4 & 24.4 $\pm$ 0.6 & 80.7 $\pm$ 0.7\\
\end{tabular}
\end{small}
\caption{Results of LBR with trivial strategies in BB/h.}\label{tab:chumps}
\end{table}

\subsection{Chumps}

In order to better understand the strengths and limitations of LBR, we first use it to evaluate simple rule-based strategies that ignore the cards completely.

First, we consider always calling, regardless of the cards. A best response (optimal counter-strategy) against this strategy is to wait until all cards are dealt and go all-in if the probability of winning is higher than 0.5. This strategy  gains on average $\frac{1}{4}$ of all players chips per hand. The best response would go all-in on half of the hands and it would win $\frac{3}{4}$ and lose $\frac{1}{4}$ of these bets. With the stack of 200 big blinds used in the ACPC, the exploitability of the always call strategy is approximately 50 BB/h.

The results of LBR are presented in Table~\ref{tab:chumps}. If we limit LBR to choose only actions fold or call (note, it has no reason to fold), both players play always call and the expected value of LBR is 0. If we include any bets, LBR always uses only the largest one against an opponent that never folds. If we use LBR in all rounds of the game, LBR gains 33 BB/h. The reason it does not achieve the best-response value is that LBR greedily bets all-in as soon as the probability of winning is higher than 0.5. In the situations in which the probability drops below the threshold until the end of the game, it loses utility compared to the actual best response. If we force LBR to call in the first two rounds and allow other bets only in rounds 3-4, this effect is minimized and LBR gains almost the whole 50 BB/h. If we use LBR only in the last round and call in the remaining 3, LBR actually gets the full 50 BB/h in expectation.

Second chump strategy we evaluate is calling with 50\% of actions and playing a random raise otherwise. Since this strategy raises often,  LBR can already gain several blinds per hand even using only fold and call. Even with this strategy, checking until more public cards are dealt increases the performance of LBR. Overall, this strategy seems to be less exploitable than always calling, since it forces LBR to fold and to commit more chips with less information about the cards.The results in Table~\ref{tab:chumps} show that LBR can already gain several blinds per hand.

The last chump strategy we evaluate is a random legal action. This strategy is most exploitable. The reason is that after the all in bet, only fold and call are legal actions. Hence, the bot will fold half of his hands in this situation, even the hands it would otherwise clearly win.

\subsection{ACPC agents}

We continue with evaluating the agents from past ACPC competitions. All these agents are based on solving a smaller abstract game by CFR and using a translation mechanism to use the strategy in the full game. The agents differ mainly in construction of the abstract game and modifications of the CFR algorithm to speed-up convergence at relevant parts of the game. The specific bots are Hyperborean 2013 and 2014 created by the University of Alberta, and the second and third best performing bots from ACPC2016. We approached the authors of all three placing submissions from the competition, but only the two were able to support this evaluation before the paper deadline. Furthermore, the winning submission used purification to meet the competition disk limit \cite{brown2016tartanian} and therefore we expect it to be highly exploitable.

\begin{table*}[t]
\centering
%\begin{small}
\begin{tabular}{c|c|r|r|r|r||r}
Betting & Rounds & Hyp. 2013 & Hyp. 2014 & Slumbot 2016 & Act1 2016 & Full Cards\\\hline
fc & 1-4 & 1048 $\pm$ ~~68 & 721 $\pm$ ~~56 & 522 $\pm$ ~~50 & 407 $\pm$ ~~47 & -424 $\pm$ 37\\
fc & 3-4 & 1006 $\pm$ ~~76 & 608 $\pm$ ~~61 & 496 $\pm$ ~~55 & 390 $\pm$ ~~55 &  -819 $\pm$ 52 \\\hline
fcpa & 3-4 & 4040 $\pm$ 147 & 3852 $\pm$ 141 & 4020 $\pm$ 115 & 2597 $\pm$ 140&  -536 $\pm$ 87\\\hline
on-tree & 3-4 & 4743 $\pm$ 163 & 4789 $\pm$ 156 &  & &  -536 $\pm$ 87\\\hline
56 bets & 1-4 & 619 $\pm$ 117 & 574 $\pm$ 125 & 763 $\pm$ ~~84 & 2429 $\pm$ 134 & 1607 $\pm$ 76 \\
56 bets & 3-4 & 5062 $\pm$ 152 & 4675 $\pm$ 152 & 3763 $\pm$ 104 & 3302 $\pm$ 122 &  2403 $\pm$ 87\\
\end{tabular}
%\end{small}
\caption{Lower bound on exploitability (in mBB/h) of ACPC bots and a bot with no card abstraction and fold-call-pot-all\_in betting computed using Local Best Respons restricted to the given betting options and to check/call out of the denoted rounds.}\label{tab:acpc}
\end{table*}

Table~\ref{tab:acpc} summarizes the results. We evaluate the players with LBR's betting restricted to just fold and call (fc); fold, call, pot, and all in (fcpa); the bets that are used by the agent in its abstract game solution (on-tree); and fold, call, all in, and 55 pot fractions computed as $0.05\cdot(1.15)^k$ for $k=0\dots54$ (56bets). If a pot fraction is not applicable, min-bet or all in is evaluated instead. The column Rounds in the table defines the rounds (or streets) in which LBR actually computed the local best response to choose actions. The remaining rounds were always check/call. All the results are averaged over 2$\times$50,000 duplicate hands. It means that each hand was played by each player on the first as well as the second position to reduce the effect of luck.

The main result is that with 97.5\% confidence, exploitability of each evaluated bot is over 3180 mBB/h. It means that folding every hand would cost the bots at least 4 times less money than playing their strategy against their worst case opponent. Second, it is important to wait when using LBR. If we use all 56 bets from the very beginning, LBR often exploits the opponent almost a full order of magnitude less than if we force it to check in the first two rounds and use the 56 bets only afterwards. The reason, as in the case of always call, is the greedy nature of LBR. It places large bets too early, without sufficiently exploiting the information it might learn later. Alternatively, it pushes the opponent to folds before she places more money in to the pot to make a larger gain. Third, all bots lose substantially even if LBR is allowed to play only fold and call. Allowing fold only later in the game does not seem to be beneficial, since the ability to exploit the information LBR learns during the hand is very limited. Most of the exploitation of LBR against all bots is realized with only using a single pot-bet option (``fcpa'' row in Table~\ref{tab:acpc}). For Hyperboreans, exploitability is further increased by adding more actions, but adding actions beyond the actions used in the abstract game (on-tree) helps only with the 2013 version and does not help at all for the 2014 version. For Slumbot and Act1, we do not know the exact betting abstraction used in their abstract game, but the betting options in ``fcpa'' are almost always included. If these bets are included, as in case of Hyperboreans, there is no need for any translation, the play will never leave the pre-computed abstract tree and all the exploitability is caused only by the errors in the card abstraction. Recent APCP bots are generally well converged within their abstract games.

The least exploitable bot in these experiments is Act1, even though it was beaten in one-on-one play by Slumbot in the competition. It confirms that as with full-game best response, even LBR may not be indicative of actual one on one performance (and vice versa).

The experiments with ACPC bots were performed on a cluster of AMD Opteron 6172 nodes with 24 cores, 32 GB of RAM and the strategies on a shared network drive. Typically, we were running 10-20 instances of LBR on each node in batches of 1000 hands. One batch for fold-call betting completed within half an hour, one batch of 56 bets experiments generally took up to 8 hours. Since an important bottle neck is the disk/network bandwidth which is further improved by caching, the variance on required resources is rather large. Still, computing good LBR values even for complicated strategies with endgame solving is perfectly feasible with the presented method.

\subsection{Full cards}

The last bot we evaluate is a large bot that uses complete non-abstracted information about the cards and the sparse ``fcpa'' betting abstraction. It plays a slightly smaller game with a 100 BB stack. The results on this bot (Table~\ref{tab:acpc}) show that LBR does not realize the actual best response and can even be substantially beaten (i.e., an uninformative lower bound approximation). When restricted to the same ``fcpa'' abstraction used in the opponent's abstraction, a full (non-local) best response shows the opponent is exploitable for 90 mBB/h, while LBR loses 536 mBB/h. However, the solution with ``fcpa'' abstraction is not sufficient to ensure low exploitability in the whole game with existing translation techniques. With hard translation used in bets off-tree, LBR wins 2403 mBB/h against this bot using 56 bets only in the last two rounds of the game. Soft translation seems to mitigate this problem a little, but definitely does not solve it.  Using sampled soft translation with 10 samples for estimating the strategy, LBR on the last two rounds is winning 1981 $\pm$ 224. The larger confidence interval is caused by playing fewer hands, since even the look-up in the huge compressed strategy without card abstraction is expensive. Note that not all 56 actions are necessary to see the high exploitability using LBR. It is sufficient to use several bets out of the original betting abstraction. For example, the bot with hard translation looses 1849 mBB/h with only fold, call, min-bet, 2, 4, 8 times pot bet, and the all in bet.

\section{Conclusions}
This paper presents the Local Best Response method for fast approximation of exploitability of large poker strategies.  If a bot is able to provide a strategy for all hands it could hold in a specific public state of the game within a reasonable time (i.e., minutes), this method can generally be used to approximate its exploitability. This is also the case for all published endgame solving techniques, which resolve a sub-game for all possible private hands at once.  

Using this method we show that the existing poker bots, including the second and the third best performing bots in the ACPC in 2016, all have exploitability substantially larger than folding all hands. The bots that use card abstraction are losing over 3 big blinds per hand on average against their worst case opponent. Exploitability can be reduced by not using card abstraction, but that necessarily leads to using a very sparse betting abstraction, which can be heavily exploited as well. Therefore, we assume that a substantial paradigm shift is necessary to create bots that would closely approximate equilibrium in full no-limit Texas hold'em.

\subsection*{Acknowledgements}
We would like to thank ACPC 2016 poker bot authors Eric Jackson and Tim Reiff for providing their bots and implementing the interface that allowed us to use LBR to evaluate them. Our tool is based on the University of Alberta Computer Poker Research Group's code base and we are grateful to all current and previous members that contributed to its development. Computing resources were provided by Calcul Quebec, Westgrid, and Compute Canada. This work was partially supported by Czech Science Foundation (15-23235S).

\begin{small}
\bibliography{references}
\bibliographystyle{aaai}
\end{small}
\end{document}